\documentclass[letter]{aa} 
\usepackage{color}

\usepackage{natbib}
\bibpunct{(}{)}{;}{a}{}{,} 
\usepackage{graphicx}
\usepackage{txfonts}
\def\kms{km s$^{-1}$}
\def\kmss{km s$^{-1}$\space}
\def\micron{$\mu$m}
\def\microns{$\mu$m\space}
\def\arcsec{$^{\prime\prime}$}
\def\arcsecs{$^{\prime\prime}$\space}

\def\deg{$^{\circ}$}
\def\degs{$^{\circ}$\space}

\def\n2h{N$_2$H$^+$}

\def\13co{$^{13}$CO}
\def\c18o{C$^{18}$O}
\def\12co{$^{12}$CO}

\def\c+{C$^+$}

\def\h2{H$_2$}

\begin{document}
   \title{[C\,{\sc ii}] 158$\mu$m  line detection of the warm ionized medium in the Scutum--Crux spiral arm tangency}

\titlerunning{[C\,{\sc ii}] Detection of the Warm Ionized Medium}
\authorrunning{Velusamy, Langer, Pineda, Goldsmith}


   \author{T. Velusamy
          \inst{1},
          W. D. Langer
           \inst{1},
           J. L. Pineda
           \inst{1},
            \and
         P. F. Goldsmith\inst{1}\fnmsep\thanks{{\it Herschel} is an ESA space observatory with science instruments provided by European-led Principal Investigator consortia and with important participation from NASA.}
          }

  \offprints{ \email{Velusamy@jpl.nasa.gov}}

   \institute{Jet Propulsion Laboratory, California Institute of Technology,
              4800 Oak Grove Drive, Pasadena, CA 91109, USA\\
              \email{Thangasamy.Velusamy@jpl.nasa.gov}
             }

   \date{Received 29/03/2012; accepted 26/04/2012}

\abstract{The \textit {Herschel} HIFI GOT C+ Galactic plane [C\,{\sc ii}] spectral survey has detected strong emission at the spiral arm tangencies.  }{ We use  the unique viewing geometry of the Scutum-Crux (S--C) tangency near \textit{ l} = 30\degs to detect the warm ionized medium (WIM) component traced by [CII] and to  study  the effects of spiral  density waves on Interstellar Medium (ISM) gas.}{We compare [C\,{\sc ii}] velocity features with ancillary H\,{\sc i}, $^{12}$CO and $^{13}$CO data near tangent velocities at each longitude to separate the cold neutral medium   and the warm neutral + ionized  components in the S--C  tangency, then we identify  [C\,{\sc ii}] emission at the highest velocities without any contribution from $^{12}$CO  clouds, as WIM.} {We present the GOT C+ results for the S--C tangency. We interpret the diffuse and extended excess [C\,{\sc ii}] emission at and above the tangent velocities as arising in the electron--dominated warm ionized gas in the WIM.  We  derive an  electron density  in the range of  0.2 -- 0.9 cm$^{-3}$ at each longitude, a factor of several higher than the average value from H$\alpha$ and pulsar dispersion. }{ We interpret the excess [C\,{\sc ii}] in S--C tangency  as shock compression of the WIM induced by the spiral  density waves.}




 \maketitle
%
\section{Introduction}
In spiral galaxies the transition from diffuse to molecular clouds, followed by star formation, and the disruption of molecular clouds and termination of star formation are greatly influenced by the interaction of the interstellar medium (ISM) with spiral density waves. The spiral tangent regions \citep[c.f.][]{vallee2008,benjamin2009} are  ideal laboratories to study the interaction of the interstellar gas and spiral density waves in the Milky Way. The tangents provide a unique viewing geometry with sufficiently long path lengths to detect and trace the diffuse atomic or ionized components of the ISM in emission from C$^+$ and N$^+$ fine-structure lines, and therefore allow us to study how the diffuse atomic and ionized components are transformed into  dense molecular clouds. The 1.9 THz [C\,{\sc ii}] line ($^2$P$_{3/2}$--$^2$P$_{1/2}$ transition of C$^+$  at 158\micron) emission in the ISM can be excited by collisions with electrons, and atomic and molecular hydrogen in regions with a  kinetic temperature T$_k$ $\gtrsim$30-40 K.  COBE FIRAS observed the strongest [C\,{\sc ii}] and [N\,{\sc ii}] emission along the Galactic spiral arm tangencies and \citet{steiman2010} fit the COBE results with four well-defined logarithmic spiral arms in the gaseous component of the ISM.  However, COBE's 7\degs beam and spectrally unresolved lines preclude obtaining detailed information on the scale and  properties of the gas in the spiral tangencies, nor whether they arise from PDRs or diffuse gas. The HIFI GOT C+ (Galactic Observations of Terahertz C+) survey of the Milky Way \citep[][]{Langer2011MW2011} also detects the strongest [C\,{\sc ii}] emission near the spiral arm tangential directions.

\begin{figure}[!hb]
\centering
\includegraphics[scale=0.35,angle=-90]{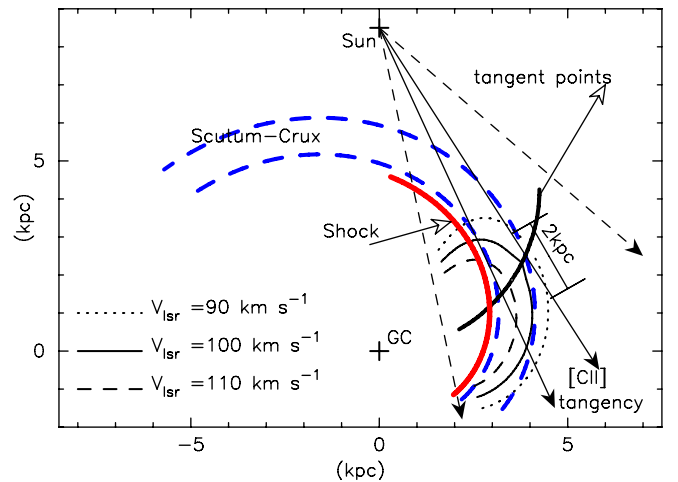}
\caption{Schematic of the Scutum--Crux spiral arm (blue dashed lines).
The dashed arrows indicate  the  GOT C+ mapped data presented in this paper, and solid arrows the tangency longitude range traced by [C\,{\sc ii}] emission. The   radial velocity   contours, spiral density wave shock  (red arc), and the loci of the tangent points (heavy dark arc) are indicated. }
\label{fig:schematic}
\end{figure}
\begin{figure*}[t]
\centering
\includegraphics[scale=0.8,angle=-90]{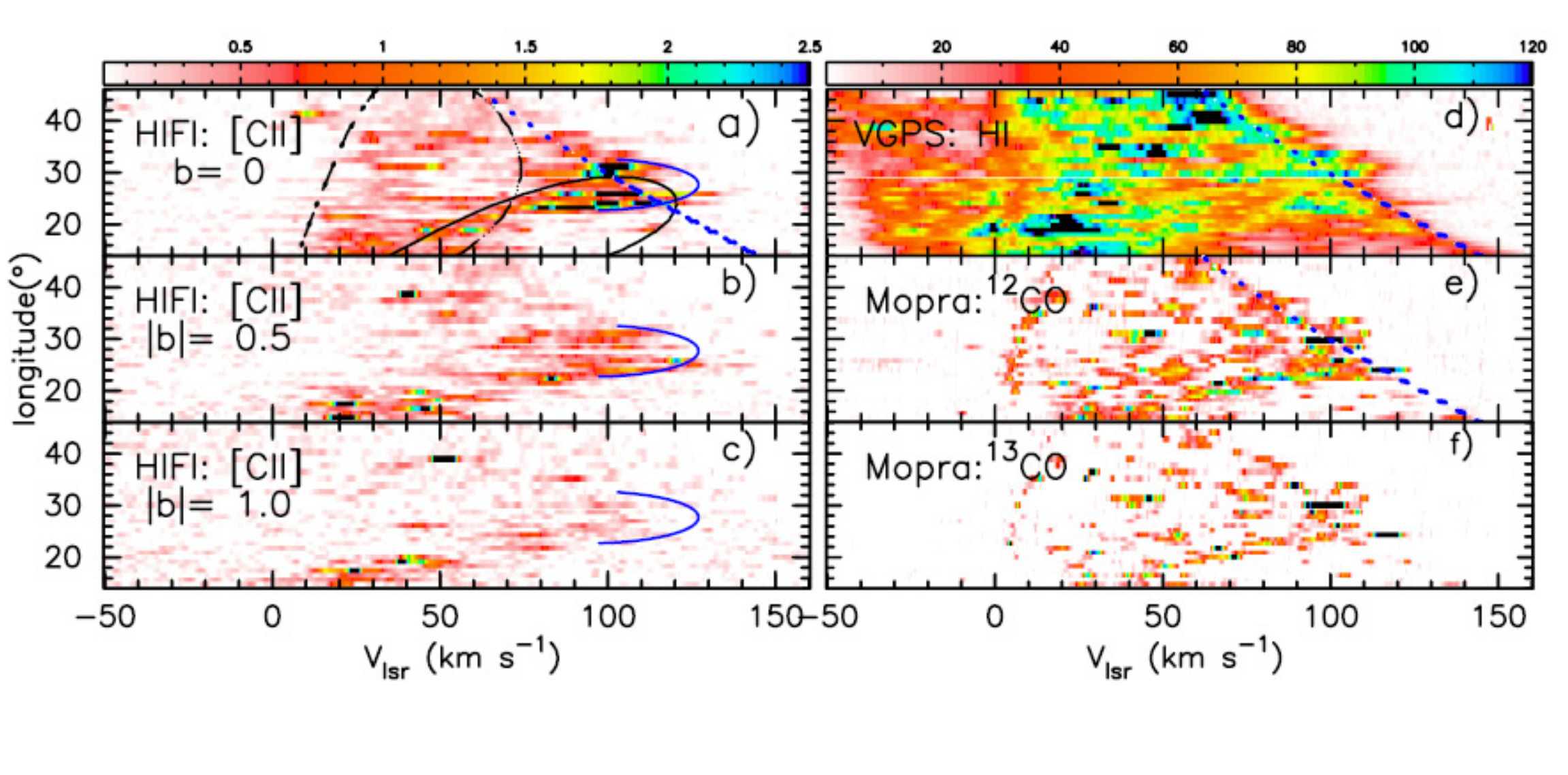}
\caption{{\it (l-v)} maps of [C\,{\sc ii}], H\,{\sc i} and CO emissions in the vicinity of the S--C tangency: (a) -- (c)  {\it (l-v)} maps of [C\,{\sc ii}] intensity at {\it b} =0.0\deg,  $\pm$0.5\degs and  $\pm$1.0\deg.   The arc (blue) indicates the loci of the velocities in the S--C tangent traced by [C\,{\sc ii}].    The loci of the spiral arm  velocities from Galactic rotation are shown in (a) as contours: Scutum--Crux (solid), Sagittarius--Carina (dashed).  (d) to (f):  {\it (l-v)} intensity maps at b=0.0\degs of  HI, $^{12}$CO(1-0) and $^{13}$CO(1-0).  The  tangent velocity as a function of longitude (see text) is shown by the dashed line (blue).  The color bars shown on top  represent the [C\,{\sc ii}]  and H\,{\sc i} map intensities, T$_{mb}$(K).    The  highest intensity values (dark blue) in the CO maps  are 15K  ($^{12}$CO) and 4K  ($^{13}$CO).}
\label{fig:l-vmap}
\end{figure*}

The GOT C+ survey consists of $\sim$500 lines--of--sight (LOS) distributed in a volume--weighted sampling of the Galactic disk. The first results based on $\sim$15 LOS  \citep[][]{langer2010,velusamy2010,pineda2010} focused on the narrow velocity [C\,{\sc ii}] features, tracing the distribution and characteristics of the  dense  atomic and molecular components of the interstellar medium, including the so-called ``dark H$_2$ gas'' (\h2 not traced by CO or H\,{\sc i}).  However, the detection of the more diffuse [C\,{\sc ii}] emission from the warm ionized medium (WIM) in the HIFI data required completion of the GOT C+ survey and enhanced data processing tools.  The WIM is a major Galactic component, with T$_{kin}$ $\sim$ 8000K, and LOS averaged density $\langle n(e)\rangle$ $\sim$ 0.03-0.10 cm$^{-3}$, with volume filling factor f$_v$ $\sim$ 0.1 in the plane and increasing to $\sim$ 0.4 at 1 kpc  above the plane \citep[see review by][based on H$\alpha$ and pulsar data]{Haffner2009}.
 The ionization and heating sources of the WIM and its relationship to the other phases of the ISM are not well understood. Carbon radio recombination lines (RRL) have been observed  in the ionized gas in H\,{\sc ii} regions \citep[][]{quireza2006} and in extended diffuse regions in the Galactic disk \citep[][]{roshi2002}.  However, detecting and characterizing the 158\microns [C\,{\sc ii}] emission in the ionized gas is critical to our understanding the Galactic WIM component and its contribution to the [C\,{\sc ii}] luminosity in galaxies.
As shown in Fig. 1 the geometry of the Scutum--Crux (S--C) spiral arm tangent is very favorable to study the structure and kinematics of the WIM component.  Here we present the [C\,{\sc ii}] emission along the S--C spiral arm tangent near Galactic longitude {\it l}=30\deg, indicating higher electron  densities,  a likely result of compression by density wave shocks.


\section{Observations}
The observations reported here are a subset of the GOT C+ Galactic plane [C\,{\sc ii}] survey at 1900.5469 GHz taken with the HIFI instrument \citep[][]{degraauw2010} on {\it Herschel}  \citep[][]{pilbratt2010}. The [C\,{\sc ii}] spectra were obtained using the Wide Band Spectrometer  (0.17 \kmss velocity resolution  over 350 \kms) in band 7b.  For each target Galactic longitude we used the Load CHOP (HPOINT) with a sky reference at {\it b} = $\pm$2\deg at each longitude. The data were processed in HIPE-8 using the standard  HIFI pipeline.  We  mitigated the electronic (HEB) and optical band 7b standing waves, following procedures in \citet{higgins2011}: (i) the HEB standing waves are removed with the HEBBaselineFit  (subtracting profile- matched off-source spectra  using a script provided by the HIFI data processing  team); (ii) the optical standing waves are dealt with  by subtracting the HEBBaselineFitted OFF-source from the ON-source spectra.  We also extracted the OFF-source spectrum to examine and correct for any [C\,{\sc ii}] spectral contamination in the OFF-source subtraction.  The $^{12}$CO and $^{13}$CO Mopra data \citep[][]{pineda2010} were observed at the [C\,{\sc ii}] LOS, and the H\,{\sc i} data were extracted from the VGPS survey \citep[][]{stil2006}.
\section{Results}

Figure 1 is a schematic of the Scutum--Crux tangency, which is a 2--D cartographic model of the Galactic spiral structure using a 4-arm spiral model \citep[c.f.][]{vallee2008,steiman2010}.  An observational challenge in tracing the distribution and properties of the interstellar gas is to separate the diffuse ionized and atomic components from the dense PDRs and molecular clouds.  To do so, we have complemented our [C\,{\sc ii}] data with observations of the H\,{\sc i} 21-cm and CO isotopologue lines to  trace the diffuse atomic and molecular gas, respectively.  These auxiliary data allow us to isolate the origin of the [C\,{\sc ii}] emission.  Fig. 2 summarizes all the GOT C+ [C\,{\sc ii}] observations and ancillary H\,{\sc i} and CO data for the inner Galaxy  containing the S--C spiral tangency.  The [C\,{\sc ii}], H\,{\sc i}, and CO emission are presented in the form of longitude--velocity {\it (l--v)} intensity maps.  These maps are sparsely sampled in longitude, every 0.9\degs for {\it l} $< $30\degs and 1.3\degs for {\it l} $>$30\deg.  For easy viewing
the intensities at each {\it l} are shown by thick bars; the true widths, corresponding  to the beam size 12\arcsec, 33\arcsecs and 60\arcsec, respectively for [C\,{\sc ii}], CO and H\,{\sc i}, are much narrower.  The S-C tangency is restricted in
longitude from 23\degs to 33\degs (Fig. 1).   In the  {\it (l-v)} maps ({\it l} = 15\degs to 45\deg)   the enhanced [C\,{\sc ii}] emission at radial velocities $>$100 \kmss (close to and beyond the tangent velocities) is very prominent at {\it l} = 23\degs to 33\deg, in striking contrast to
other longitudes.  It is the strongest in the plane  and easily traced at {\it b} = $\pm$0.5\deg,  but marginal at {\it b} = $\pm$1.0\degs (Fig. 2 a--c).
\begin{figure*}[t]
\centering
\includegraphics[scale=0.85,angle=-90]{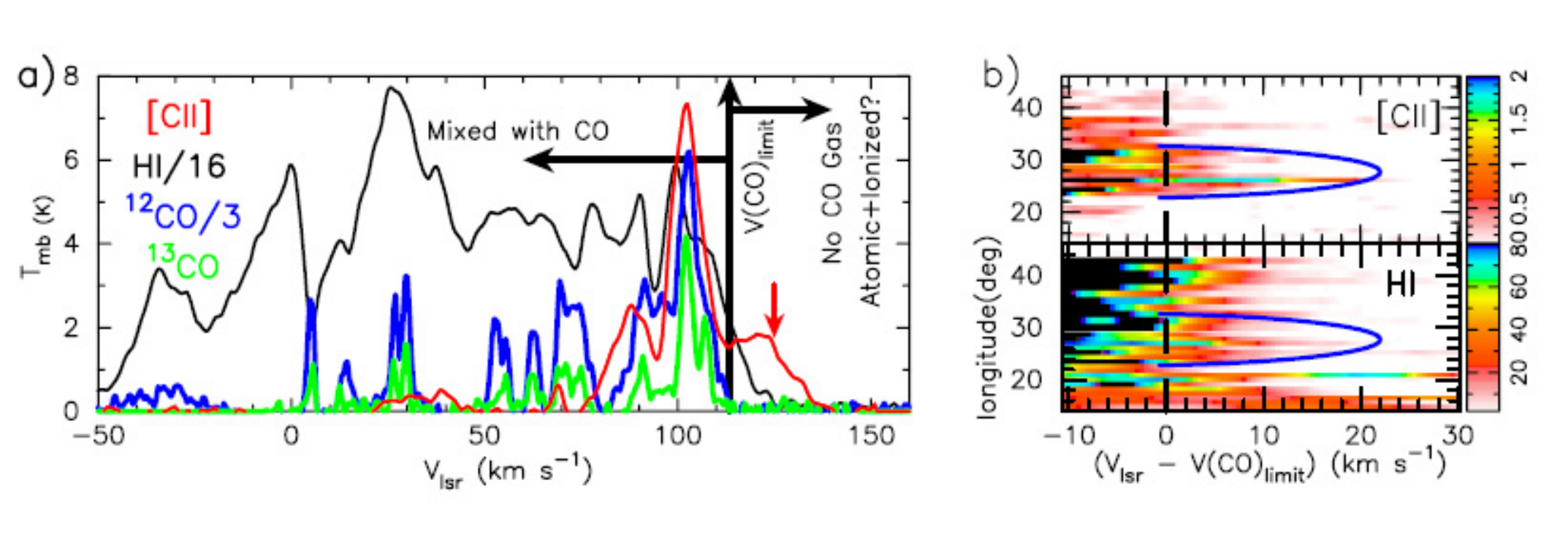}
\caption{(a) [C\,{\sc ii}]  H\,{\sc i}, and CO spectra for LOS: G026.1+0.0,
in the S--C tangency.  The arrows indicate the velocities of the gases with and without CO (black) and the  [C\,{\sc ii}] excess (red).
(b) {\it (l-v)} maps of [C\,{\sc ii}] and H\,{\sc i} intensities as function of  V$_{lsr}$-V(CO)$_{limit}$.     The color bars shown on the right represent the map intensities, T$_{mb}$(K ). The arcs (blue) delineate the longitudes and velocities of the S-C tangency traced by  [C\,{\sc ii}].
 }
\label{fig:spectra}
\end{figure*}

The overlay of the radial velocities V$_{lsr}$ on the [C\,{\sc ii}] {\it (l--v)} map in Fig. 2a brings out the importance of [C\,{\sc ii}] as a tracer of spiral arms in the inner Galaxy.  The Scutum--Crux (S-C) arm, which is not that well traced by other tracers in these longitudes \citep[c.f.][]{vallee2008}, is well traced by the [C\,{\sc ii}] emission.   However the Sagittarius--Carina arm is only partially traced.  This   difference between the S--C and the outer arms is primarily because [C\,{\sc ii}] preferably traces the high FUV environments within Galactic radius $\sim$6 kpc 
(Pineda et al, in prep.).  The loci of the tangent velocities at each longitude (broken line in Figs. 2a, d, and e) were calculated for the Galactic rotation parameters for the first quadrant  from H\,{\sc i} by \citet[hereafter LHB]{levine2008}.  The tangent velocities are model dependent, nevertheless, near the S--C tangency both H\,{\sc i} and [C\,{\sc ii}]  show emission significantly beyond the tangent velocities,  well above the velocities of the $^{12}$CO clouds (Figs. 2, 3a).    Thus,  we can use the CO velocity data to separate the [C\,{\sc ii}] associated with discrete  molecular clouds   from that with the warm  diffuse atomic and ionized gas  components as shown in Fig. 3a.  In this [C\,{\sc ii}] spectrum we note (red arrow) bright anomalous [C\,{\sc ii}] emission at velocities greater than the   tangent velocity ($\sim$110 \kms), which is very distinct from other [C\,{\sc ii}] spectral features, as it is associated with  weak H\,{\sc i}  while also having no CO counterpart.

To bring out more clearly the exceptional characteristics of the [C\,{\sc ii}] emission at the S--C tangency velocities on a larger scale we show in Fig. 3b  {\it (l--v)} maps of the highest velocity [C\,{\sc ii}] and H\,{\sc i} components without any CO.
To make this display  we define V(CO)$_{limit}$ (in Fig. 3a) for each longitude, as the highest velocity at which discrete $^{12}$CO clouds are detected. Then we reconstruct the spectra at each longitude with V$_{lsr}$-V(CO)$_{limit}$ as the velocity axis.  Therefore, at each longitude emission at V$_{lsr}$-V(CO)$_{limit}$ $>$0 traces the ISM with no CO clouds and the [C\,{\sc ii}] and H\,{\sc i} emission arising in the diffuse atomic/ionized gas.  Although at V$_{lsr}$-V(CO)$_{limit}$ $>$0 H\,{\sc i} emission is seen at all longitudes, the [C\,{\sc ii}] is detected only in the longitudes near the S--C tangency with the exception of  {\it l}=36.4\degs (Fig. 3b). We conclude that the relative intensities of [C\,{\sc ii}], H\,{\sc i} and CO emissions are remarkably different for the S-C tangency compared to those for other velocities or longitudes with relatively little H\,{\sc i} and no CO.  Thus we have isolated a diffuse atomic and/or ionized gas component at the highest velocity along the S-C arm.
Assuming circular Galactic rotation, V$_{lsr}$ velocities for the [C\,{\sc ii}] emission indicate this ionized gas is located in the S-C arm at radial distances 3.6 to 4.2 kpc from the Galactic center across the longitude range {\it l} = 23\degs to 33\degs (using r$_{sun}$=8.5 kpc).
\section{Discussion}
The anomalous excess [C\,{\sc ii}] emission in the S--C spiral arm tangency is the first example of the direct  detection of the large scale Galactic diffuse ionized gas (WIM) by the 158\microns [C\,{\sc ii}] line emission.   From the COBE data only indirect estimates of the fraction of [C\,{\sc ii}] intensity arising from the WIM could be made by using the [NII] intensities \citep[][]{steiman2010} due to its poor spatial and spectral resolution.  Our {\it Herschel} HIFI detection of the diffuse ionized gas provides detailed spatial and kinematic information on the nature of this gas component in spiral arms.  A possible explanation for this [C\,{\sc ii}] emission is that it originates from ionized gas that has been compressed, or is being compressed, by the spiral density wave shocks (Fig. 1).  We locate the ionized gas traced by [C\,{\sc ii}] in the spiral tangency using the kinematic distances derived from the observed radial velocities (V$_{lsr}$).  The schematic shown in Fig. 1 demonstrates  that the velocity contours derived from the LHB rotation curve for V$_{lsr}$  = 90, 100, and 110 \kmss can be used to  constrain a linear path length (L)   for this large-scale diffuse emission across each longitude.   For  {\it l} $>$ 30\degs we assume L $\sim$1 kpc, as these observed radial velocities have no solution in  Galactic rotation models.  In Table 1 we list the range of radial velocities over which [C\,{\sc ii}] is detected across the S--C tangency and the inferred path lengths.

We can use the H\,{\sc i} and [C\,{\sc ii}] integrated intensities along each LOS (Table 1) to estimate the line of sight averaged densities,  $\langle n(H)\rangle$  and $\langle n(e)\rangle$ assuming warm neutral (WNM)  and ionized (WIM) conditions,  respectively. Defining  $\langle n(H)\rangle = N(H)/L$, where $N(H)=1.82\times10^{18}I(HI)$ cm$^{-2}$ and $L$ is the path length along the spiral arm tangent, we find  $\langle n(H)\rangle$  $\sim$ 0.05 - 0.21 cm$^{-3}$ (Table 1) for a range of $N(H)$ = 3.3--5.9$\times$ 10$^{20}$ cm$^{-2}$.  For the WNM and WIM conditions (T$_{k}$ =8000K; \citet{wolfire2003}) the critical density for excitation of [C\,{\sc ii}] by H atoms is $\sim$1300 cm$^{-3}$, and for electrons $\sim$ 45 cm$^{-3}$.
We show below that the required $\langle n(H)\rangle$ to produce the [C\,{\sc ii}] emission from a WNM is too large compared to the density estimates from H\,{\sc i} observations.  Following \citet{langer2010},   assuming that this WNM component  is uniformly distributed along the tangency, we  estimate the minimum required $\langle n(H)\rangle$ $\sim$ 1.7 - 7.1 cm$^{-3}$  to produce the $I(CII)$ in Table 1 using the corresponding path lengths for each LOS. These densities are factors of 20 to 150 larger than the $\langle n(H)\rangle$ inferred from H\,{\sc i} intensities and column densities.
Furthermore, in the {\it (l-v)} maps in Fig 3b, we detect significant H\,{\sc i} emission at V$_{lsr}$ $>$ V(CO)$_{limit}$ at all longitudes, whereas [C\,{\sc ii}] emission is restricted to longitudes of the S--C tangency. If the [C\,{\sc ii}] emission were excited only by H collisions we would expect to detect it at all longitudes in proportion to $I(HI)$, which is not observed.   Finally, we  analyze the possibility of \h2 excitation and find  $\langle$n(\h2)$\rangle$ $\sim$ 5 - 17 cm$^{-3}$ is needed  to produce the observed $I(CII)$.    Such low \h2 densities over  long path lengths are  unrealistic for \h2 formation, especially, given the low $\langle n(H)\rangle$. The observed broad velocity ranges for the [C\,{\sc ii}] emission over 10\degs longitude also precludes it originating in  compact high density regions.  Clearly the [C\,{\sc ii}] emission is not produced  by H or \h2 excitation.   Furthermore, in the low resolution {\it (l-v)} maps of hydrogen RRL, the S--C tangency is evident at these velocities \citep[][]{alves2012}.

In the diffuse medium the excitation is sub-thermal and the emission is optically thin, therefore the intensity in an ionized gas is given by, $I(CII)=\int T_A(CII)dv$ $= 6.9\times10^{-16} n(e)/n_{cr}(e)\,exp(-91.3/T) N(C^+)\,(K km s^{-1})$, where $n_{cr}(e)$ is the critical de--excitation density.  Assuming constant density and temperature along the LOS and fully ionized carbon, we have $N(C^+)=n(C^+)L$.  We can substitute $n(C^+) = x(C^+)n_t$, where  $x(C^+$) is the fractional abundance of carbon, and $n_t=n(H^+)+n(H)+2n(H_2)$. Inserting $n(e)=x(e)n_t$, and  $n_{cr}(e)=21.4T_3$$^{0.37}$, where $T_3$=$T/1000$, into the equation for $I(CII)$  yields a solution for $n_t=0.32T_3^{0.18}[I(CII)/(x(e)x_{-4}(C^+)L_{kpc})]^{0.5}$, where x$_{-4}$ is in units of 10$^{-4}$ and $L_{kpc}$ is in kpc. In a fully ionized gas $x(e)$=1 and  adopting $x(C^+)$ =2.9$\times$10$^{-4}$ in the S-C tangency (appropriate for $R_{Gal}$ $\sim$4 kpc  \citep[][]{wolfire2003}), yields
$\langle n_t \rangle \sim 0.19T_3^{0.18}(I(CII)/L_{kpc})^{0.5}$.
\begin{table}[!t]
\caption{[C\,{\sc ii}] \& H\,{\sc i} intensities and derived parameters for S-C tangency}
\renewcommand{\tabcolsep}{0.05cm}
\begin{tabular}  {c c c c c c c c}
\hline\hline
LOS	&	V$_{lsr}$	&	$\Delta$V	&	$L$$^\dag$	&	$I(CII)$	&	$I(HI)$	&	$\langle n(HI)\rangle$	&	 $\langle n(e)\rangle$$^\ddag$	 \\
 	&	\kms	&	\kms	&	kpc	&	K\kms	&	K\kms	&	cm$^{-3}$	&	cm$^{-3}$	\\
\hline
G23.5+0.0	&	116	&	6	&	0.6	&	1.8	&	217	&	0.21	&	 0.45	\\
G24.3+0.0	&	131	&	12	&	0.8	&	2.9	&	186	&	0.14	&	 0.50	\\
G25.2+0.0	&	118	&	12	&	2.7	&	5.7	&	262	&	0.06	&	 0.38	\\
G26.1+0.0	&	125	&	24	&	2.3	&	26.5	&	184	&	0.05	&  0.90	\\
G27.0+0.0	&	114	&	18	&	2.9	&	4.9	&	548	&	0.11	&	 0.34	\\
G27.8+0.0	&	109	&	7	&	2.6	&	1.8	&	222	&	0.05	&	 0.22	\\
G28.7+0.0	&	117	&	12	&	1.8	&	4.0	&	141	&	0.05	&	 0.39	\\
G30.0+0.0	&	112	&	12	&	1	&	3.1	&	210	&	0.12	&	 0.46	\\
G31.3+0.0	&	119	&	12	&	1	&	4.4	&	325	&	0.19	&	 0.55	\\
G32.6+0.0	&	112	&	12	&	1	&	2.7	&	291	&	0.17	&	 0.44	\\
average b=0.0	&	117	&	12	&	1.8	&	6.7	&	530	&	0.17	&	 0.51	\\
average b=0.5	&	107	&	11	&	2.2	&	3.3	&	740	&	0.20	&	 0.32	\\
average b=1.0	&	93	&	10	&	3.5	&	1.5	&	672	&	0.12	&	 0.17	\\
\hline
\end{tabular}\\
\footnotetext{} { $^\dag$ \textit{Constrained by the highest velocity contours in Fig.1}}\\
 \footnotetext{} {$^\ddag$ \textit{For $x(e)$=1 and   $x(C^+)$ =2.9$\times$10$^{-4}$ }}\\
\end{table}

Using this approach, assuming a fully ionized gas, x(e)=1,  we calculate $\langle n(e)\rangle$ = $\langle n_t \rangle$ for all LOS (Table 1)
for T$_{k}$ = 8000K.  These estimates, assuming uniform conditions, are reasonable considering that f$_v$ $>$ 0.1 in the plane implies a WIM LOS filling factor $\sim$~0.5  \citep[][]{Haffner2009}. These values are strictly a lower limit if the gas is partially ionized, $x(e)<$1, but only weakly so, as $\langle n_t\rangle  \propto x(e)^{-0.5}$.   The electron densities estimated from [C\,{\sc ii}],  $\langle n(e)\rangle$  in  Table 1 are in the range of 0.2 -- 0.9 cm$^{-3}$ with a mean  $\sim$ 0.46 cm$^{-3}$ and 1$\sigma$ scatter of 0.2 cm$^{-3}$.
We can rule out the excess [C\,{\sc ii}] emission in the S-C tangency as H\,{\sc ii} regions and their photoionized envelopes. Although, in the \textit {l-v} map of   H\,{\sc ii} regions  there are about 10 within the longitude range 23\deg--33\degs of the S-C tangency
at velocities above 110 \kmss  \citep[][and references therein]{anderson2009}, none overlaps the observed [C\,{\sc ii}]  LOS (Table 1).  Furthermore, Anderson et al. report that in this region only 20\% of the H\,{\sc ii} regions do not have detections in $^{13}$CO, which is  likely to be even less for $^{12}$CO detections.  Thus by excluding all the velocities having detectible $^{12}$CO (see Section 3), in effect we have excluded the H\,{\sc ii}  regions while identifying the [C\,{\sc ii}] excess as representing the WIM gas.

The average  densities from the [C\,{\sc ii}] excess,  in Table 1,  are
several times higher than the LOS averaged densities inferred from pulsar dispersion and H$\alpha$ measurements, n$_e$ $\sim$0.08 cm$^{-3}$  \citep[][]{Haffner2009} and we argue that our larger mean value is a result of compression by the WIM-spiral arm interaction.  The relatively low H\,{\sc i} emission also implies that we are detecting the WIM and not the WNM. Though local variations by a factor 5 over the global mean density of 0.08 cm$^{-3}$ may be present in the WIM, here, in the S-C tangency we detect  these high electron  densities  all along the spiral arm,  over a long path length ($>$ 1 kpc).  Furthermore,
that such elevated densities  (by  factors $\sim$3 --11) are detected only  at the spiral arm tangency and not beyond is evidence that this enhancement is unique to this location and viewing geometry (Fig. 1). It is, therefore, conceivable that  here we are detecting  the WIM which is being compressed by its interaction with the gravitational well of the spiral arms and is on the path to becoming WNM.
The [C\,{\sc ii}] emission extends beyond the V(CO)$_{limit}$ typically to $\sim$12 -18 \kmss
indicative of the presence of large non-circular velocity components due to shocks in the gas traced by [C\,{\sc ii}].  This scenario for the shock induced compression and velocities is consistent with the interstellar gas response to spiral pattern speed as seen in hydrodynamical models \citep[][]{martos2004jkas}.  The lack of a smooth distribution in longitude of the WIM all across the S-C tangency as traced by [C\,{\sc ii}] (Fig. 3b) is possibly the result of a spatially complex gaseous response to an orderly spiral potential
\citep[see][]{yanez2008}. We also estimate $\langle n(e)\rangle$ averaged over the tangency at {\it b}=0.0\deg, $\pm$0.5\degs (z=70pc) and $\pm$1.0\degs (z=140pc)
 by averaging over all {\it l} (from 23\degs to 33\deg) and  integrating over 10 \kmss above the V(CO)$_{limit}$.   The  results (Table 1), in spite of the uncertainties in the path lengths, seem to indicate a trend in the decrease of $\langle n(e)\rangle$ with increasing z,  while $\langle n(H)\rangle$ is roughly constant. Such a trend would be consistent with the spiral density  shocks being more effective near the plane of the Galaxy.

 In this paper we present the first detection of the WIM in  spectrally resolved  [C\,{\sc ii}]  158 \microns emission. Future \textit {Herschel}/SOFIA [NII] spectral data will help characterize the WIM.
 Our interpretation of  [C\,{\sc ii}] excess  tracing the WIM in the S--C tangency is further supported by similar detections along other spiral tangencies of S--C  at {\it l} $\sim$310\degs and Norma-3kpc including the start of Perseus  at {\it l} = 330\deg--342\degs    \citep[][]{Langer2011MW2011}.   It is conceivable  that the WIM traced by [C\,{\sc ii}] at the S--C tangency exists all around the spiral arms, but unlike the tangent direction it is not easy to detect towards other directions due to insufficient path length and velocities blended with other components. Thus  it  may add significantly to the total [C\,{\sc ii}] luminosity in galaxies.

\begin{acknowledgements}
We acknowledge the helpful comments by the referee. We thank the staffs of the ESA and NASA Herschel Science Centers for their help. HIFI has been designed and built by a consortium
of institutes and university departments from across Europe, Canada
and the United States under the leadership of SRON Netherlands Institute
for Space Research Groningen, The Netherlands and with major contributions
from Germany, France, and the US. Consortium members are: Canada:
CSA, U.Waterloo; France: CESR, LAB, LERMA, IRAM; Germany: KOSMA,
MPIfR, MPS; Ireland: NUI Maynooth; Italy: ASI, IFSI-INAF, Osservatorio
Astrofisico di Arcetri-INAF; Netherlands: SRON, TUD; Poland: CAMK, CBK;
Spain: Observatorio Astronomico Nacional (IGN), Centro de Astrobiologia
(CSIC-INT); Sweden: Chalmers University of Technology – MC2, RSS \&
GARD, Onsala Space Observatory, Swedish National Space Board, Stockholm
University – Stockholm Obseratory; Switzerland: ETH Zurich, FHNW; USA:
Caltech, JPL, NHSC. This work was performed at the Jet Propulsion Laboratory, California Institute of Technology, under contract with the National Aeronautics and Space Administration.
\end{acknowledgements}

\bibliographystyle{aa}
\bibliography{ms}

\begin{thebibliography}{20}
\expandafter\ifx\csname natexlab\endcsname\relax\def\natexlab#1{#1}\fi

\bibitem[{{Alves} {et~al.}(2012){Alves}, {Davies}, {Dickinson}, {Calabretta},
  {Davis}, \& {Staveley-Smith}}]{alves2012}
{Alves}, M.~I.~R., {Davies}, R.~D., {Dickinson}, C., {et~al.} 2012, \mnras, (in
  press)

\bibitem[{{Anderson} {et~al.}(2009){Anderson}, {Bania}, {Jackson}, {Clemens},
  {Heyer}, {Simon}, {Shah}, \& {Rathborne}}]{anderson2009}
{Anderson}, L.~D., {Bania}, T.~M., {Jackson}, J.~M., {et~al.} 2009, \apjs, 181,
  255

\bibitem[{{Benjamin}(2009)}]{benjamin2009}
{Benjamin}, R.~A. 2009, in IAU Symposium 254, 319--322

\bibitem[{{de Graauw} {et~al.}(2010){de Graauw}, {Helmich}, {Phillips},
  {Stutzki}, {Caux}, {Whyborn}, {Dieleman}, {Roelfsema}, {Aarts}, {Assendorp},
  {Bachiller}, {Baechtold}, {Barcia}, {Beintema}, {Belitsky}, {Benz}, {Bieber},
  {Boogert}, {Borys}, {Bumble}, {Ca{\"i}s}, {Caris}, {Cerulli-Irelli},
  {Chattopadhyay}, {Cherednichenko}, {Ciechanowicz}, {Coeur-Joly}, {Comito},
  {Cros}, {de Jonge}, {de Lange}, {Delforges}, {Delorme}, {den Boggende},
  {Desbat}, {Diez-Gonz{\'a}lez}, {di Giorgio}, {Dubbeldam}, {Edwards},
  {Eggens}, {Erickson}, {Evers}, {Fich}, {Finn}, {Franke}, {Gaier}, {Gal},
  {Gao}, {Gallego}, {Gauffre}, {Gill}, {Glenz}, {Golstein}, {Goulooze},
  {Gunsing}, {G{\"u}sten}, {Hartogh}, {Hatch}, {Higgins}, {Honingh}, {Huisman},
  {Jackson}, {Jacobs}, {Jacobs}, {Jarchow}, {Javadi}, {Jellema}, {Justen},
  {Karpov}, {Kasemann}, {Kawamura}, {Keizer}, {Kester}, {Klapwijk}, {Klein},
  {Kollberg}, {Kooi}, {Kooiman}, {Kopf}, {Krause}, {Krieg}, {Kramer},
  {Kruizenga}, {Kuhn}, {Laauwen}, {Lai}, {Larsson}, {Leduc}, {Leinz}, {Lin},
  {Liseau}, {Liu}, {Loose}, {L{\'o}pez-Fernandez}, {Lord}, {Luinge}, {Marston},
  {Mart{\'{\i}}n-Pintado}, {Maestrini}, {Maiwald}, {McCoey}, {Mehdi}, {Megej},
  {Melchior}, {Meinsma}, {Merkel}, {Michalska}, {Monstein}, {Moratschke},
  {Morris}, {Muller}, {Murphy}, {Naber}, {Natale}, {Nowosielski}, {Nuzzolo},
  {Olberg}, {Olbrich}, {Orfei}, {Orleanski}, {Ossenkopf}, {Peacock}, {Pearson},
  {Peron}, {Phillip-May}, {Piazzo}, {Planesas}, {Rataj}, {Ravera}, {Risacher},
  {Salez}, {Samoska}, {Saraceno}, {Schieder}, {Schlecht}, {Schl{\"o}der},
  {Schm{\"u}lling}, {Schultz}, {Schuster}, {Siebertz}, {Smit}, {Szczerba},
  {Shipman}, {Steinmetz}, {Stern}, {Stokroos}, {Teipen}, {Teyssier}, {Tils},
  {Trappe}, {van Baaren}, {van Leeuwen}, {van de Stadt}, {Visser}, {Wildeman},
  {Wafelbakker}, {Ward}, {Wesselius}, {Wild}, {Wulff}, {Wunsch}, {Tielens},
  {Zaal}, {Zirath}, {Zmuidzinas}, \& {Zwart}}]{degraauw2010}
{de Graauw}, T., {Helmich}, F.~P., {Phillips}, T.~G., {et~al.} 2010, \aap, 518,
  L6

\bibitem[{{Haffner} {et~al.}(2009){Haffner}, {Dettmar}, {Beckman}, {Wood},
  {Slavin}, {Giammanco}, {Madsen}, {Zurita}, \& {Reynolds}}]{Haffner2009}
{Haffner}, L.~M., {Dettmar}, R.-J., {Beckman}, {et~al.} 2009, Rev. Mod. Phys,
  81, 969

\bibitem[{{Higgins}(2011)}]{higgins2011}
{Higgins}, D.~R. 2011, {Advanced Optical Calibration of the Herschel HIFI
  Heterodyne Spectrometer, PhD Thesis} (National Univ. Ireland Maynooth.)

\bibitem[{{Langer}(2011)}]{Langer2011MW2011}
{Langer}, W.~D. 2011, in MilkyWay2011, http://mw2011.ifsi-roma.inaf.it/

\bibitem[{{Langer} {et~al.}(2010){Langer}, {Velusamy}, {Pineda}, {Goldsmith},
  {Li}, \& {Yorke}}]{langer2010}
{Langer}, W.~D., {Velusamy}, T., {Pineda}, J.~L., {et~al.} 2010, \aap, 521, L17

\bibitem[{{Levine} {et~al.}(2008){Levine}, {Heiles}, \& {Blitz}}]{levine2008}
{Levine}, E.~S., {Heiles}, C., \& {Blitz}, L. 2008, \apj, 679, 1288

\bibitem[{{Martos} {et~al.}(2004){Martos}, {Ya{\~n}ez}, {Hernandez}, {Moreno},
  {Moreno}, {Moreno}, \& {Pichardo}}]{martos2004jkas}
{Martos}, M., {Ya{\~n}ez}, M., {Hernandez}, X., {et~al.} 2004, J Korean Astr
  Soc, 37, 199

\bibitem[{{Pilbratt} {et~al.}(2010){Pilbratt}, {Riedinger}, {Passvogel},
  {Crone}, {Doyle}, {Gageur}, {Heras}, {Jewell}, {Metcalfe}, {Ott}, \&
  {Schmidt}}]{pilbratt2010}
{Pilbratt}, G.~L., {Riedinger}, J.~R., {Passvogel}, T., {et~al.} 2010, \aap,
  518, L1

\bibitem[{{Pineda} {et~al.}(2010){Pineda}, {Velusamy}, {Langer}, {Goldsmith},
  {Li}, \& {Yorke}}]{pineda2010}
{Pineda}, J.~L., {Velusamy}, T., {Langer}, W.~D., {et~al.} 2010, \aap, 521, L19

\bibitem[{{Quireza} {et~al.}(2006){Quireza}, {Rood}, {Balser}, \&
  {Bania}}]{quireza2006}
{Quireza}, C., {Rood}, R.~T., {Balser}, D.~S., \& {Bania}, T.~M. 2006, \apjs,
  165, 338

\bibitem[{{Roshi} {et~al.}(2002){Roshi}, {Kantharia}, \&
  {Anantharamaiah}}]{roshi2002}
{Roshi}, D.~A., {Kantharia}, N.~G., \& {Anantharamaiah}, K.~R. 2002, \aap, 391,
  1097

\bibitem[{{Steiman-Cameron} {et~al.}(2010){Steiman-Cameron}, {Wolfire}, \&
  {Hollenbach}}]{steiman2010}
{Steiman-Cameron}, T.~Y., {Wolfire}, M., \& {Hollenbach}, D. 2010, \apj, 722,
  1460

\bibitem[{{Stil} {et~al.}(2006){Stil}, {Taylor}, {Dickey}, {Kavars}, {Martin},
  {Rothwell}, {Boothroyd}, {Lockman}, \& {McClure-Griffiths}}]{stil2006}
{Stil}, J.~M., {Taylor}, A.~R., {Dickey}, J.~M., {et~al.} 2006, \aj, 132, 1158

\bibitem[{{Vall{\'e}e}(2008)}]{vallee2008}
{Vall{\'e}e}, J.~P. 2008, \aj, 135, 1301

\bibitem[{{Velusamy} {et~al.}(2010){Velusamy}, {Langer}, {Pineda}, {Goldsmith},
  {Li}, \& {Yorke}}]{velusamy2010}
{Velusamy}, T., {Langer}, W.~D., {Pineda}, J.~L., {et~al.} 2010, \aap, 521, L18

\bibitem[{{Wolfire} {et~al.}(2003){Wolfire}, {McKee}, {Hollenbach}, {Tielens},
  {Tielens}, \& {Tielens}}]{wolfire2003}
{Wolfire}, M.~G., {McKee}, C.~F., {Hollenbach}, D., {et~al.} 2003, \apj, 587,
  278

\bibitem[{{Y{\'a}{\~n}ez} {et~al.}(2008){Y{\'a}{\~n}ez}, {Norman}, {Martos},
  {Hayes}, {Hayes}, \& {Hayes}}]{yanez2008}
{Y{\'a}{\~n}ez}, M.~A., {Norman}, M.~L., {Martos}, M.~A., {et~al.} 2008, \apj,
  672, 207

\end{thebibliography}
\end{document}